\documentclass[11pt]{article}
\usepackage{amsmath,amssymb}
\usepackage{framed}
\usepackage{color}

\usepackage{graphicx}

\usepackage{amsmath,latexsym}

\begin{document}

\setcounter{page}{1}

\pagestyle{plain}

\begin{center}
\Large{\bf A Tachyon Field around the Black Hole}\\
\small \vspace{1cm} {Narges
Rashidi}$^{a,b}$\footnote{n.rashidi@umz.ac.ir (Corresponding Author)}  \\
\vspace{0.5cm} $^{a}$ Department of Theoretical Physics, Faculty of Science,
University of Mazandaran,\\
P. O. Box 47416-95447, Babolsar, IRAN\\
\vspace{0.5cm} $^{b}$ ICRANet-Mazandaran,
University of Mazandaran,\\
P. O. Box 47416-95447, Babolsar, IRAN\\
\end{center}

\begin{abstract}
We study the effects of the presence of the tachyon field around the
black hole. We show that in presence of the tachyon field, unlike
the ordinary canonical scalar field, the time evolution of the black
hole mass depends on the potential of this field. By considering
several types of potential, we study the behavior of the black hole
mass and its time evolution and find some interesting results. We
find that the presence of the tachyon field causes the accretion of
the mass into the black hole. We also show that with linear and
hilltop potentials, in some ranges of the parameters space, the mass
of the black hole can decrease even without any Hawking radiation.
\end{abstract}
\newpage

\section{Introduction}\label{sec1}
Given that in the current universe there are many black holes that
have different masses, the study of black holes, as the
thermodynamical systems, has attracted a lot of attention. As
regards some massive black holes have originated from primordial
black holes, one of the interesting aspects in the study of the
black holes is the accretion of the matter into the black hole.
Although the standard argument assumes that the mass of the
primordial black holes remains constant once they are
formed~\cite{Carr74,Bet81,Cus98}, the authors of Ref.~\cite{Bea02}
have shown that this is not necessarily true. They have studied
scenarios in which the accretion of the black holes' mass occurs.
Satisfying the null energy condition causes that the mass of the
black hole never decreases in a classical way, but even
increases~\cite{Haw73}. However, by considering the quantum
processes, there would be the Hawking radiation leading to the mass
decrement of the black hole~\cite{Haw74,Haw75}. By emitting the
Hawking radiation, the black holes can be evaporated completely. So,
any black hole in its evolution has been affected by both decrement
(arising from the Hawking radiation) and increment (due to the
accretion of the matter-energy) of the mass. If the Hawking
radiation dominates, the black hole evaporates. If the accretion
dominates, the black hole grows.

In Ref.~\cite{Bea02}, it is discussed that the primordial black
holes might have grown by absorbing the existing scalar field in the
universe, leading to the mass increment of the black holes. Also, in
Refs.~\cite{Bab04,Bab05} it has been shown that by considering the
phantom scalar field, the mass of the black hole effectively
decreases. In Ref.~\cite{Rod09}, the authors have considered the
nonminimally coupled canonical scalar field and studied the
evolution of the black hole mass. They have shown that the accretion
of the nonminimally coupled scalar field causes the mass to decrease
when the black hole mass is smaller than a certain critical value.
Note that, this result came up without exist the phantom energy in
the model.

Another interesting scalar field that can be responsible for both
the initial and late time acceleration phase of the universe is the
tachyon
field~\cite{Sen99,Sen02a,Sen02b,Noz13,Noz18,Ras18,Ras20,Ras21}. In
this paper, instead of considering the phantom field or non-minimal
coupling, we adopt the tachyon field to study the accretion of the
black hole. We show that when we consider the tachyon field in the
theory, the time derivative of the black hole mass would be
corresponding to the potential of this field. Note that, in the
minimal ``simple canonical scalar field'', there is no dependence on
the potential in the time derivative of the black hole
mass~\cite{Bea02}. In this respect, in our model, it is possible to
study the accretion of the mass in the minimally coupled scalar
field with different potentials. On the other hand, the authors of
Refs~\cite{Fel02,Mc02,Sin18} have shown that a negative potential
can lead to interesting cosmological results. In this regard, we
show that by adopting the suitable negative potential, there would
be a decrease in the black hole mass, at least in some ranges of the
model's parameter space. This result is obtained without presenting
a nonminimal coupling. Note that, the negative potential for the
tachyon field leads to the negative energy density. Although
currently there is no evidence for the role of the negative energy
density in our universe, some realizations of the relativistic
quantum field theories predict this possibility~\cite{Kuo97,For09}.
Also, some authors have studied the models with negative energy
density and have shown that, at least at a theoretical level, it
leads to interesting cosmological implications~\cite{Pro11,Nem14}.
Therefore, it seems inspiring to consider both negative and positive
potentials for the tachyon field.

With these preliminaries, the paper is organized as follows. In
section 2, we consider the tachyon field around the black hole. By
using the energy-momentum tensor of the tachyon field and also its
equation of motion, we seek the effects of this field on the mass
evolution of the black hole. In this section, we obtain an
expression for the time evolution of the black hole mass in terms of
the tachyon field's potential. In section 3, we study the evolution
of the black hole mass by considering several types of potential. We
show that by considering the tachyon field around the black hole, it
is possible to have accretion of the mass into the black hole. We
also show that, with linear and hilltop potentials, the mass of the
black hole can decrease without considering the Hawking radiation.
In section 4, we summarize our work.

\section{Tachyon Field around the Black Hole} \label{sec2}

With a tachyon field, we deal with the following action
\begin{eqnarray}
\label{eq1} S=\int
d^{4}x\sqrt{-g}\Bigg[\frac{1}{2\kappa^{2}}R-V(\phi)\sqrt{1-2
	X} \Bigg],
\end{eqnarray}
where $R$ is the Ricci scalar, $\kappa$ is the gravitational
constant, and $V(\phi)$ is the potential of the tachyon field. Also,
$X$ is defined as
$X=-\frac{1}{2}g^{\mu\nu}\partial_{\mu}\phi\,\partial_{\nu}\phi$.It
should be noticed that, as it has been demonstrated in
Ref.~\cite{Ran14}, if we consider
$X=\frac{1}{2}g^{\mu\nu}\partial_{\mu}\phi\,\partial_{\nu}\phi$ the
tachyon field becomes a phantom. However, in this work we don't
consider this case. Note that, in Ref.~\cite{Bea02}, where a
canonical scalar field has been considered, the authors have studied
both zero and non-zero potentials. However, in the case with a
tachyon field, considering the zero potential removes the effect of
the tachyon field entirely.

The equation of motion of the tachyon field, obtained from the
action (\ref{eq1}), is given by
\begin{eqnarray}
\label{eq2}\Box\phi-\partial_{\mu}\phi\,\partial^{\mu}\phi\frac{
	\Box
	\phi}{1+\partial_{\mu}\phi\,\partial^{\mu}\phi}=\frac{V'}{
	V}\,.
\end{eqnarray}
Also, we have the following energy-momentum tensor, corresponding to
the tachyon field
\begin{eqnarray}
\label{eq3}T_{\mu\nu}=\frac{V}{\sqrt{1+\partial^{\nu}\phi\,\partial_{\nu}\phi}}\partial_{\nu}\phi\partial_{\nu}\phi
-g_{\mu\nu}\,V\,\sqrt{1+\partial^{\nu}\phi\,\partial_{\nu}\phi}\,.
\end{eqnarray}

The interesting point is that, although our considered tachyon field
is not a usual phantom, depending on the values of the potential, it
is possible for the tachyon field to behave like a phantom. In fact,
from equation (\ref{eq3}) we have $\rho+p=-\frac{VX}{\sqrt{1-2X}}$,
where $\rho=T_{00}$ and $p=T_{ii}$ with $i=1,2,3$ (see also
Ref.~\cite{Bab07}, where it has been shown that it is possible to
express density and pressure in terms of Lagrangian and its
derivatives). Considering that the parameter $X$ is given by
$X=\frac{1}{2}\dot{\phi}^{2}$, the tachyon field violates the null
energy condition for positive values of the potential. In this way,
although the tachyon field is different from the usual phantom
field, for the positive values of the potential, the tachyon field
has the ``phantom-like'' behavior. On the other hand, for the
negative values of the potential, we have $\rho+p>0$, meaning the
tachyon field doesn't violate the null energy condition. While, in
this case, the weak energy condition is violated (since we have
$\rho<0$), and in some realizations of the relativistic quantum
field theories this violation is predicted and possible.

Now, we consider the Schwarzschild coordinates and obtain the
spherically symmetrical version of (\ref{eq2}) as
\begin{eqnarray}
\label{eq4}\ddot{\phi}-\left(1-\frac{2M}{r}\right)\frac{1}{r^{2}}\partial_{r}\left[\left(1-\frac{2M}{r}\right)r^{2}\partial_{r}\phi\right]
\hspace{4cm}\nonumber\\=-\frac{V'}{V}\,\left(1-\frac{2M}{r}\right)\,
\left[1-\left(1-\frac{2M}{r}\right)^{-1}\dot{\phi}^{2}+\left(1-\frac{2M}{r}\right)(\partial_{r}\phi)^{2}\right]\,,
\end{eqnarray}
where a dot shows the derivative of the parameter with respect to
the time. We should find spherically symmetric solutions to the
equation (\ref{eq4}) with the following boundary condition according
to the Bondi accretion process,
\begin{eqnarray}
\label{eq5}\phi(t,r\rightarrow \infty)=\phi_{c}(t)\,,
\end{eqnarray}
where $\phi_{c}(t)$ shows the cosmological evolution of the scalar
field. In fact, we assume the black hole lies on a local
asymptotically flat space with boundary condition (\ref{eq5}). This
is because, in comparison with the cosmological length or time
scales, the black hole is very small~\cite{Jac99}. After we find a
solution $\phi(r,t)$ for equation (\ref{eq4}), satisfying the
boundary condition, we can obtain its energy flux through the
horizon that is completely absorbed by the black hole. In this
regard, in Schwarzschild coordinates, the rate of accretion of the
tachyon field onto the black hole is given by
\begin{eqnarray}
\label{eq6}\dot{M}=\oint_{r=2M} r^{2}\,\,T_{t}^{\,\,r}\,d\Omega\,.
\end{eqnarray}
If we adopt Eddington-Finkelstein coordinates $(v,r)$, which is
regular on the horizon, the solution of equation (\ref{eq4}) in
stationary configuration is given by
\begin{eqnarray}
\label{eq7}\phi(v,r)=C_{1}+C_{2}\left[v-r+2M\log
\Big(\frac{2M}{r}\Big)\right]\,,
\end{eqnarray}
where $C_{1}$ and $C_{2}$ are constant. In~\cite{Akh09}, the authors
have found a specific form of lagrangian leading to the stationary
solutions. This means that they have obtained a form of lagrangian
for which the action is invariant under shift symmetry
$\phi\longrightarrow \phi+\lambda$. However, with the lagrangian
form given in (\ref{eq1}), we follow a quasi-stationary
approach~\cite{Fro03}. In the quasi-stationary approach, we consider
a slowly rolling scalar field with $V'(\phi)\approx 0$ and
$\ddot{\phi}\approx 0$ which resemble the slow-roll conditions in
inflation. Although the equation (\ref{eq4}) is apparently different
from the corresponding ones in Refs.~\cite{Rod09,Fro03}, we can
consider the right-hand side of that equation as an effective term
and write
\begin{eqnarray}
\label{eq8}\ddot{\phi}-\left(1-\frac{2M}{r}\right)\frac{1}{r^{2}}\partial_{r}\left[\left(1-\frac{2M}{r}\right)r^{2}\partial_{r}\phi\right]
=-V_{eff}'\,,
\end{eqnarray}
where
\begin{eqnarray}
\label{eq9}V_{eff}'=\frac{V'}{V}\,\left(1-\frac{2M}{r}\right)\,
\left[1-\left(1-\frac{2M}{r}\right)^{-1}\dot{\phi}^{2}+\left(1-\frac{2M}{r}\right)(\partial_{r}\phi)^{2}\right]\,.
\end{eqnarray}
Note that, in equation (\ref{eq8}) and forthcoming equations, a dot
on the parameter means the derivative over the Eddington-Finkelstein
coordinate $v$ which is equal to derivative over time.

As emphasized in Ref.~\cite{Fro03}, the solution of equation
(\ref{eq8}) is rather independent of the form of the potential. In
this regard, the field configuration at the black hole horizon can
be approximated as~\cite{Fro03,Rod09,Rod12}
\begin{eqnarray}
\label{eq10}\phi(v,r)=\phi_{c}\left(v-r+2M\log
\Big(\frac{2M}{r}\Big)\right)\,,
\end{eqnarray}
where $\phi_{c}$ has been introduced in boundary condition
(\ref{eq5}). Now, by substituting the solution (\ref{eq10}) in the
equation (\ref{eq8}), we get
\begin{eqnarray}
\label{eq11}\left[1+\frac{2M}{r}+\left(\frac{2M}{r}\right)^2+\left(\frac{2M}{r}\right)^3\right]\ddot{\phi}=-V_{eff}'\,,
\end{eqnarray}
where we now have the following expression for $V_{eff}'$
\begin{eqnarray}
\label{eq12}V_{eff}'=\frac{V'}{V}\,
\left[1-\left(1-\frac{16M^{4}}{r^{4}}\right)\dot{\phi}^{2}\right]\,.
\end{eqnarray}
Note that, since we have considered the slowly varying tachyon
field, $\dot{\phi}$ is almost a constant, and the approximated
solution is valid as long as $\ddot{\phi}$ and $V'_{eff}$ are very
close to zero.

To study the time evolution of the black hole mass, we should find
$T_{t}^{\,\,r}$. From equations (\ref{eq3}) and (\ref{eq10}) we
obtain
\begin{eqnarray}
\label{eq13}T_{t}^{\,\,r}=\frac{\left(\frac{2M}{r}\right)^{2}\,V\,\dot{\phi}_{c}^{2}}{\sqrt{1-
		\left[1+\frac{2M}{r}+\left(\frac{2M}{r}\right)^2+\left(\frac{2M}{r}\right)^3\right]\dot{\phi}_{c}^{2}}}\,.
\end{eqnarray}
The important point about equation (\ref{eq13}) is the presence of
the potential. When there is a canonical scalar field, in both
minimally or nonminimally coupled cases, the potential is absent in
$T_{t}^{\,\,r}$. However, if we consider the tachyon field, the
potential is presented in $T_{t}^{\,\,r}$. This is an interesting
issue. The reason is that since $T_{t}^{\,\,r}$ is potential
dependent, by considering just the several types of potential we can
study the time evolution of the black hole mass.

In a quasi-stationary approach, it is possible to parameterize the
field absorbed at the black hole horizon, by the field approximated
from infinity as~\cite{Jac99,Rod09,Rod12}
\begin{eqnarray}
\label{eq14} \phi_{c}(t)\approx\phi_{_{\infty}}+
\dot{\phi}_{_{\infty}}(t-t_{0})\,,
\end{eqnarray}
with $\phi_{_{\infty}}$ and $\dot{\phi}_{_{\infty}}$ to be constant
parameters. In fact, in this regard, we consider $r\gg 2M$ limit and
approximate the field from the infinity to the black hole horizon.
Now, from equations (\ref{eq6}), (\ref{eq13}) and (\ref{eq14}) we
get
\begin{eqnarray}
\label{eq15}
\dot{M}=\frac{16\pi\,M^2\,V\,\dot{\phi}_{_{\infty}}^{2}}{\sqrt{1-
		\left[1+\frac{2M}{r}+\left(\frac{2M}{r}\right)^2+\left(\frac{2M}{r}\right)^3\right]\dot{\phi}_{_{\infty}}^{2}}}\,.
\end{eqnarray}
Since we consider $r\gg 2M$ limit, for small values of
$\dot{\phi}_{_{\infty}}$, we can rewrite equation (\ref{eq15}) as
\begin{eqnarray}
\label{eq16}
\dot{M}=\Bigg[16\pi\,M^2\,V\,\dot{\phi}_{_{\infty}}^{2}\Bigg]\Bigg[1+\frac{1}{2}
\bigg[1+\frac{2M}{r}+\left(\frac{2M}{r}\right)^2+\left(\frac{2M}{r}\right)^3\bigg]\dot{\phi}_{_{\infty}}^{2}\Bigg]\,.
\end{eqnarray}
Also, in the following, we just keep the terms in $\dot{M}$ which
are of the order of $M^{2}$. This means that we keep the two first
terms of the second bracket of the equation (\ref{eq16}). In the
following, using equation (\ref{eq16}), we study the evolution of
the black hole mass.

\section{Evolution of the Black Hole Mass}
To find $M$ from equation (\ref{eq16}), we follow Ref.~\cite{Rod09}
and consider $\dot{M}$ as
\begin{eqnarray}
\label{eq17} \dot{M}=f(t)\,M^{2}\,,
\end{eqnarray}
where, in our case,
\begin{eqnarray}
\label{eq18} f(t)= 16\pi\,V\,\dot{\phi}_{_{\infty}}^{2}
\bigg(1+\frac{1}{2}\dot{\phi}_{_{\infty}}^{2}\bigg)\,.
\end{eqnarray}
Note that, $f(t)$ in equation (\ref{eq18}) depends on the potential.
However, by considering the canonical scalar field, even the
non-minimally coupled one, there is no dependence on the
potential~\cite{Bea02,Rod09}. From equations (\ref{eq17}) and
(\ref{eq18}) we see that a negative potential can cause decreasing
in the black hole mass even in the absence of Hawking radiation.
This is an interesting issue and we discuss it in the next
subsections.

Now, the solution of equation (\ref{eq17}) is given by
\begin{eqnarray}
\label{eq19} M(t)=\frac{M_{0}}{1-M_{0}\,{\cal{F}}(t)}\,,
\end{eqnarray}
where $M_{0}$ is the mass at time $t_{0}$ and ${\cal{F}}$ is defined
as
\begin{eqnarray}
\label{eq20} {\cal{F}}(t)=\int_{t_{0}}^{t} f(t')\,dt'\,.
\end{eqnarray}
Since the mass is positive, so there is always a constraint on
equation (\ref{eq19}) as $M_{0}\,{\cal{F}}(t)<1$. In this step, we
should specify the form of the potential. We consider four types of
potential as linear, quadratic, natural, and hilltop potentials. In
the following, we study each case separately.

\subsection{Linear Potential}
The first potential which we consider is the linear potential as
$V\sim \phi$~\cite{Mc10}. By this potential, we find the function
$f(t)$ as
\begin{eqnarray}
\label{eq21} f(t)=8\,\dot{\phi}^{2}_{_{\infty}} \left(
\dot{\phi}^{2}_{_{\infty}} +2 \right) \pi \left(
\dot{\phi}_{_{\infty}} \, \left( t-{\it t_0} \right)
+\phi_{_{\infty}} \right)\,,
\end{eqnarray}
and ${\cal{F}}(t)$ as
\begin{eqnarray}
\label{eq22} {\cal{F}}(t)=16\,\pi\dot{\phi}^{2}_{_{\infty}} \left(
1+\frac{1}{2}\,\dot{\phi}^{2}_{_{\infty}} \right) \left(
\dot{\phi}_{_{\infty}}\, \left( \frac{1}{2}\,{t}^{2}-t_{0}\,t
\right) +\phi_{_{\infty}}\,t \right)\,.
\end{eqnarray}
By substituting equations (\ref{eq21}) and (\ref{eq22}) in equation
(\ref{eq17}) and (\ref{eq19}), we can obtain the black hole mass and
its time evolution as follows
\begin{eqnarray}
\label{eq23} M=\frac { M_{0}}{-16\,\pi \dot{\phi}_{\infty}^{2}
	\left( 1+\frac{1}{2}\,{\dot{\phi}}_{\infty}^{ 2} \right)
	\left( \dot{\phi}_{\infty}\, \left( \frac{1}{2}\,{t}^{2}-t_{0}\,t
	\right) + \phi_{\infty}\,t \right) M_{0}+1}\,,\nonumber\\
\end{eqnarray}

\begin{eqnarray}
\label{eq24} \dot{M}=\frac {16\,\pi \left( \dot{\phi}_{\infty}\,
	\left( t-t_{0} \right) +\phi_{\infty} \right) {\dot{\phi}}_{\infty}^{2} \left( 1+\frac{1}{2}\,{\dot{\phi}}_{\infty}^{2}
	\right) M_{0}^{2}}{ \left( -16\pi{\dot{\phi}}_{\infty}^{2} \left(
	1+\frac{1}{2}{\dot{\phi}}_{\infty}^{2} \right) \left(
	\dot{\phi}_{\infty} \left( \frac{1}{2}\,{t}^{2}-t_{0}\,t \right)
	+ \phi_{\infty}\,t \right) M_{0}+1 \right) ^{2}}\,.\nonumber\\
\end{eqnarray}
Equation (\ref{eq23}) sets the following constraint on the model's
parameter space in the linear potential case
\begin{eqnarray}
\label{eq25} 16\,\pi \dot{\phi}_{\infty}^{2} \left(
1+\frac{1}{2}\,{\dot{\phi}}_{\infty}^{ 2} \right)  \left(
\dot{\phi}_{\infty}\, \left( \frac{1}{2}\,{t}^{2}-t_{0}\,t \right) +
\phi_{\infty}\,t \right) M_{0}<1\,.
\end{eqnarray}

\begin{figure}[]
	\begin{center}
		\includegraphics[scale=0.5]{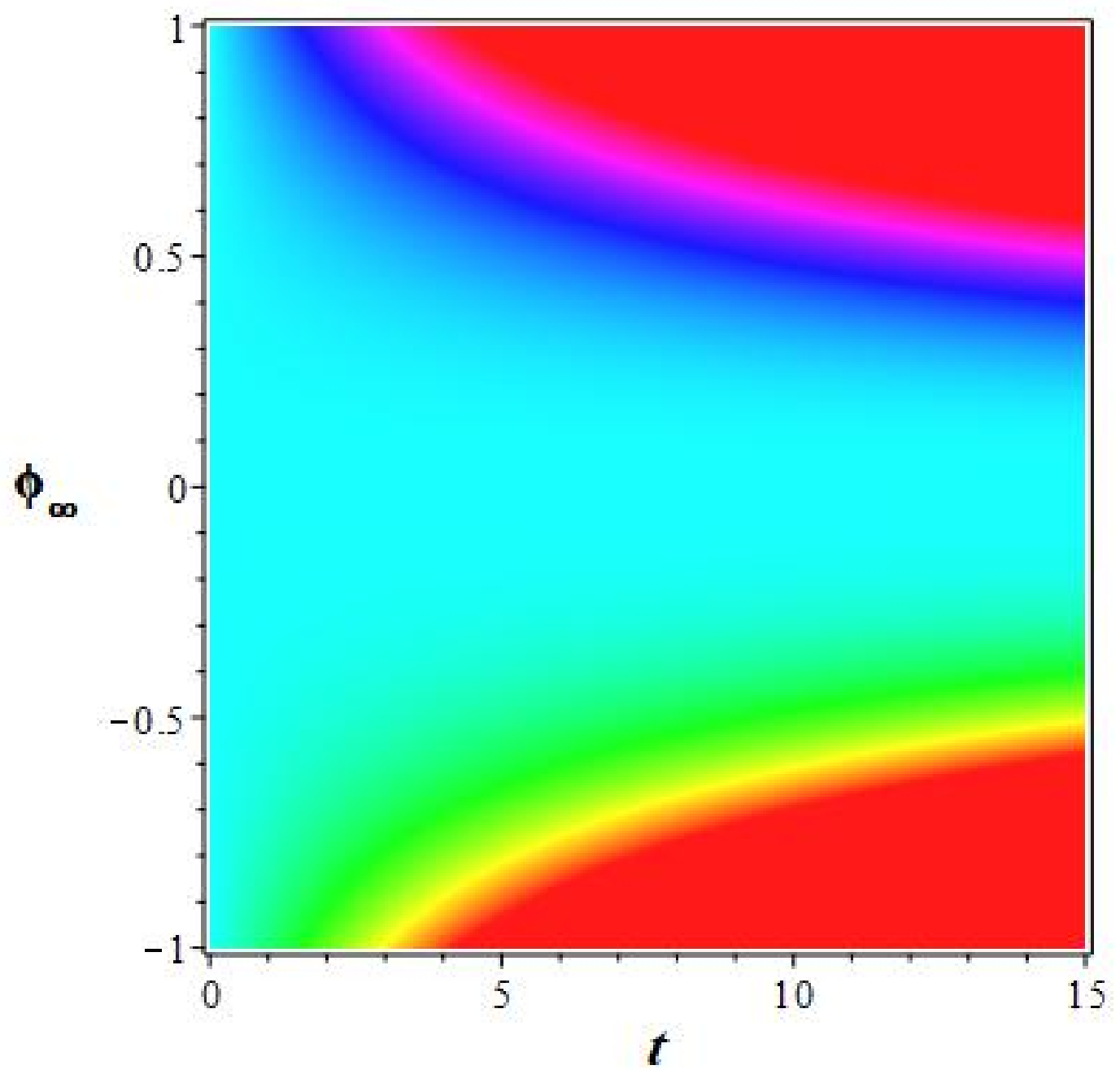}
		\includegraphics[scale=0.57]{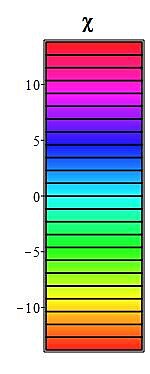}
		\includegraphics[scale=0.5]{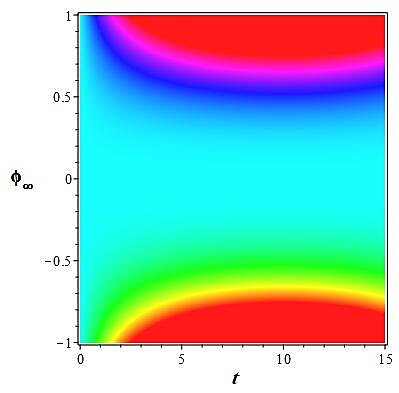}
		\includegraphics[scale=0.55]{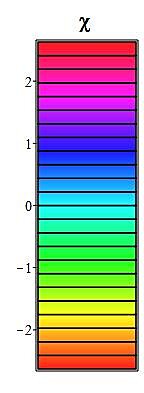}
	\end{center}
	\caption{\small {Presentation of the range of parameters
			$\phi$ and $t$, for the linear potential, satisfying the constraint
			(\ref{eq25}). The upper panel is corresponding to the case with
			$\phi_{\infty}=0.1\,\dot{\phi}_{\infty}$ and the lower panel is
			corresponding to the case with $\phi_{\infty}=-0.1\,\dot{\phi}_{\infty}$. Also, the parameter $\chi$
			represents the left-hand side of equation (\ref{eq25}).}}
	\label{fig1}
\end{figure}

Therefore, for the numerical study of the model, we should consider
the above constraint. In this regard, we perform numerical analysis
on this constraint to find the suitable ranges of the parameters. To
perform this analysis, we should specify the parameters
$\phi_{\infty}$ and $\dot{\phi}_{\infty}$. Given that these
parameters are arbitrary constant, we consider two cases as
$\phi_{\infty}=0.1\,\dot{\phi}_{\infty}$ and
$\phi_{\infty}=-0.1\,\dot{\phi}_{\infty}$. We also adopt $M_{0}=1$
and $t_{0}=0$. By these adoptions of the parameters, we perform
numerical analysis on the parameter space of $\phi$ and $t$,
satisfying the constraint (\ref{eq25}). The results are shown in
figure 1, where the parameter $\chi$ represents the left-hand side
of the constraint (\ref{eq25}). The region of $\phi_{\infty}$ and
$t$ leading to $\chi<1$ are viable. This figure helps us to adopt
appropriate values of parameter $\phi_{\infty}$ and therefore
$\dot{\phi}_{\infty}$, to study the black hole mass and its time
evolution numerically. In figure 2, we plot the evolution of the
black hole masse for $\phi_{\infty}=0.1\,\dot{\phi}_{\infty}$ and
$\phi_{\infty}=-0.1\,\dot{\phi}_{\infty}$, for some sample values of
$\phi_{\infty}$ as $0.3$, $0.1$, $-0.1$ and $-0.3$. These adopted
values of $\phi_{\infty}$ satisfy the constraint (\ref{eq25}). From
figure 2, we see that by adopting
$\phi_{\infty}=0.1\,\dot{\phi}_{\infty}$, $\phi_{\infty}=0.1$ and
$\phi_{\infty}=0.3$, we get accretion of the tachyon field into the
black hole and the black hole mass increases. By adopting
$\phi_{\infty}=0.1\,\dot{\phi}_{\infty}$, $\phi_{\infty}=-0.1$ and
$\phi_{\infty}=-0.3$, the black hole mass decreases even without
considering the Hawking radiation. In the case with
$\phi_{\infty}=-0.1\,\dot{\phi}_{\infty}$, for all adopted values of
$\phi_{\infty}$, the tachyon field at first causes increasing of the
black hole mass, then it decreases the mass. This interesting result
is obtained because of the fact that with
$\phi_{\infty}=0.1\,\dot{\phi}_{\infty}$, for $\phi_{\infty}=-0.1$
and $\phi_{\infty}=-0.3$, and with
$\phi_{\infty}=-0.1\,\dot{\phi}_{\infty}$, for all adopted values of
$\phi$, the potential can be negative leading to negative values of
$\dot{M}$. The evolution of $\dot{M}$ versus time is shown in figure
3, for both cases with $\phi_{\infty}=0.1\,\dot{\phi}_{\infty}$ and
$\phi_{\infty}=-0.1\,\dot{\phi}_{\infty}$. We conclude that, by
considering the tachyon field with the linear potential, it is
possible to have a decrease in the black hole mass without
considering the Hawking radiation. Note that, in the ranges of the
parameters space leading to black hole mass decrease, we have
$\rho+p>0$. This means that, in these ranges, the null energy
condition is not violated. Therefore, in these ranges, the tachyon
field is not a phantom. In this case, we face the mass decrease of
the black hole because of the negative energy density of the tachyon
field.

\begin{figure}[]
	\begin{center}
		\includegraphics[scale=0.35]{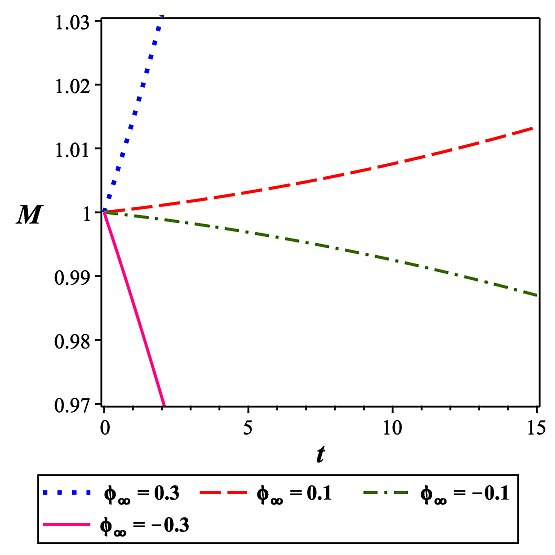}
		\includegraphics[scale=0.35]{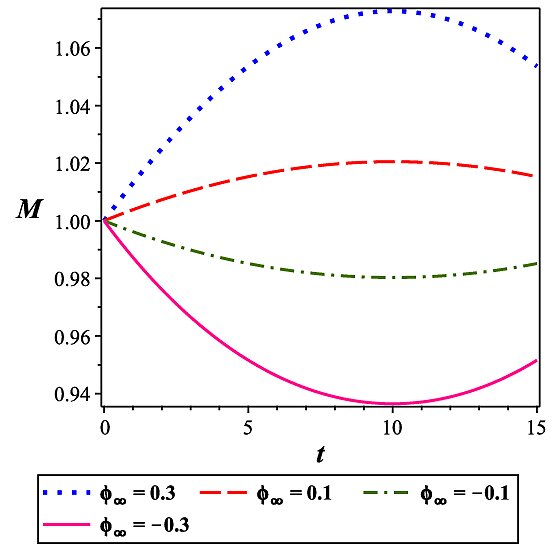}
	\end{center}
	\caption{\small {Time evolution of the black hole mass
			for different values of $\phi_{\infty}$, with linear potential. The
			left panel is corresponding to the case with $\phi_{\infty}=0.1\,\dot{\phi}_{\infty}$ and the right panel is corresponding
			to the case with $\phi_{\infty}=-0.1\,\dot{\phi}_{\infty}$.}}
	\label{fig2}
\end{figure}

\begin{figure}[]
	\begin{center}
		\includegraphics[scale=0.35]{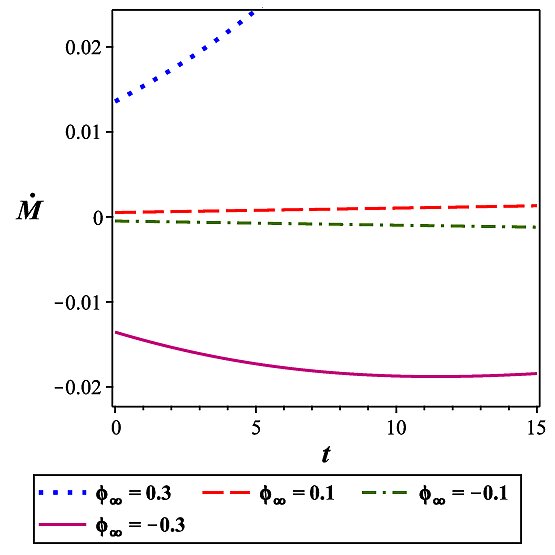}
		\includegraphics[scale=0.35]{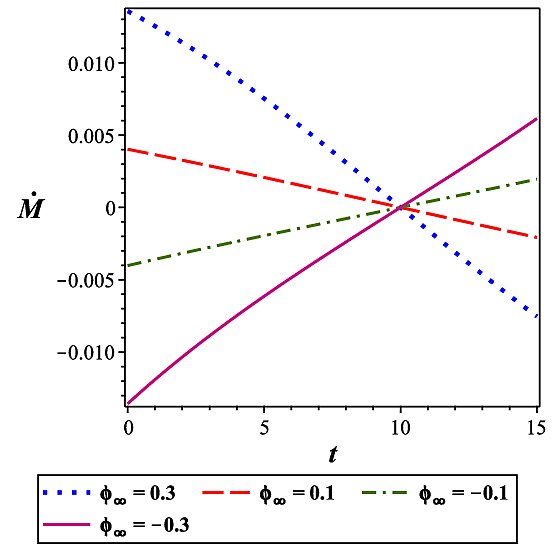}
	\end{center}
	\caption{\small {Time evolution of $\dot{M}$ for
			different values of $\phi_{\infty}$, with linear potential. The
			left panel is corresponding to the case with $\phi_{\infty}=0.1\,\dot{\phi}_{\infty}$ and the right panel is corresponding
			to the case with $\phi_{\infty}=-0.1\,\dot{\phi}_{\infty}$.}}
	\label{fig3}
\end{figure}

\subsection{Quadratic Potential}\label{sec3}
The next potential which we study is the quadratic potential as
$V\sim \phi^{2}$. With the quadratic potential, we find the
following expressions
\begin{eqnarray}
\label{eq26} f(t)=8\,\pi \left( \dot{\phi}_{_{\infty}}^{2}+2 \right)
\left( \dot{\phi}_{_{\infty}}\, \left( t- t_0 \right)
+\phi_{_{\infty}} \right) ^{2}\,\dot{\phi}_{_{\infty}}^{2}\,,
\end{eqnarray}
and
\begin{eqnarray}
\label{eq27} {\cal{F}}=\frac{16}{3}\,\pi\dot{\phi}_{_{\infty}}\,
\left( 1+\frac{1}{2}\,\dot{\phi}_{_{\infty}}^{2} \right) \bigg(
\dot{\phi}_{_{\infty}}\,t-\dot{\phi}_{_{\infty}}\,
t_0+\phi_{_{\infty}} \bigg) ^{3}\,.
\end{eqnarray}
With these expressions and by using equations (\ref{eq17}) and
(\ref{eq19}), we get
\begin{eqnarray}
\label{eq28} M=\frac {\,M_{0}}{3-8\,\pi\, \left(  \left( t-t_{0}
	\right) \dot{\phi}_{_{\infty}}+ \phi_{_{\infty}} \right) ^{3}\dot{\phi}\, \left(
	{\dot{\phi}}_{_{\infty}}^{2}+2 \right)  M_{0}} \,,
\end{eqnarray}

\begin{eqnarray}
\label{eq29} \dot{M}=\frac {72 \left( {\dot{\phi}}_{_{\infty}}^{2}+2
	\right) {\dot{\phi}}_{_{\infty}}^{2}\pi \left(  \left( t-t_{0} \right)
	\dot{\phi}_{_{\infty}}+\phi_{_{\infty}} \right) ^{2}{ M_{0}} ^{2}}{ \left( -3+8\,
	\pi\,\left( \left( t- t_{0} \right) \dot{\phi}_{_{\infty}}+\phi_{_{\infty}} \right)
	^{3}\dot{\phi}_{_{\infty}}\, \left( {\dot{\phi}}_{_{\infty}}^{2}+2 \right) M_{0}
	\right) ^{2}} \,.
\end{eqnarray}
To have the positive black hole mass, from equation (\ref{eq28}),
the following constraint should be satisfied
\begin{eqnarray}
\label{eq30} 8\,\pi\, \left(  \left( t-t_{0} \right)
\dot{\phi}_{_{\infty}}+ \phi_{_{\infty}} \right)
^{3}\dot{\phi}_{_{\infty}}\, \left( {\dot{\phi}}_{_{\infty}}^{2}+2
\right)  M_{0}<3\,.
\end{eqnarray}

\begin{figure}[]
	\begin{center}
		\includegraphics[scale=0.5]{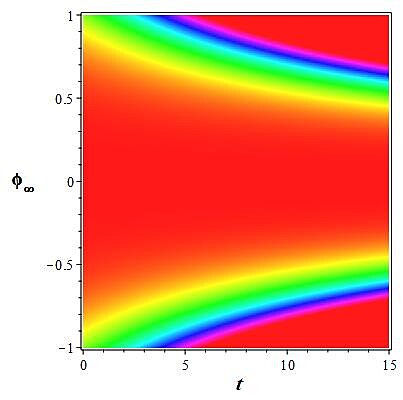}
		\includegraphics[scale=0.57]{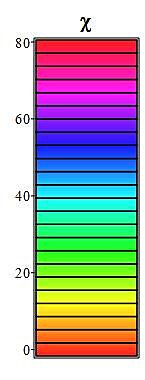}
		\includegraphics[scale=0.5]{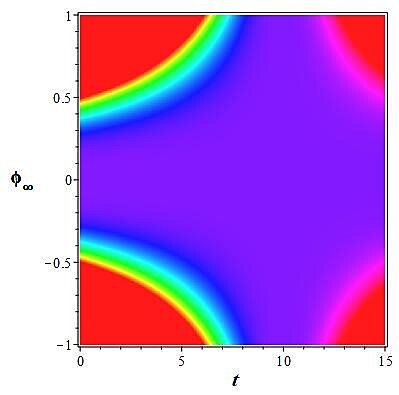}
		\includegraphics[scale=0.56]{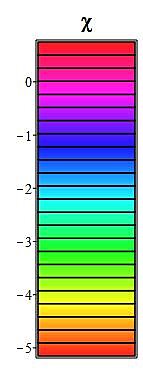}
	\end{center}
	\caption{\small {Presentation of the range of parameter
				$\phi_{\infty}$ and $t$, for the quadratic potential, satisfying
				the constraint (\ref{eq30}). The left panel is corresponding to the
				case with $\phi_{\infty}=0.1\,\dot{\phi}_{\infty}$ and the right
				panel is corresponding to the case with $\phi_{\infty}=-0.1\,\dot{\phi}_{\infty}$. Also, the parameter $\chi$
				represents the left hand side of equation (\ref{eq30}).}}
	\label{fig4}
\end{figure}

\begin{figure}[]
	\begin{center}
		\includegraphics[scale=0.35]{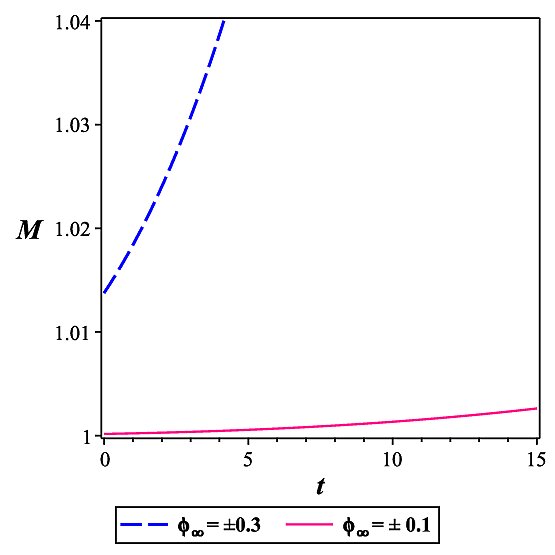}
		\includegraphics[scale=0.35]{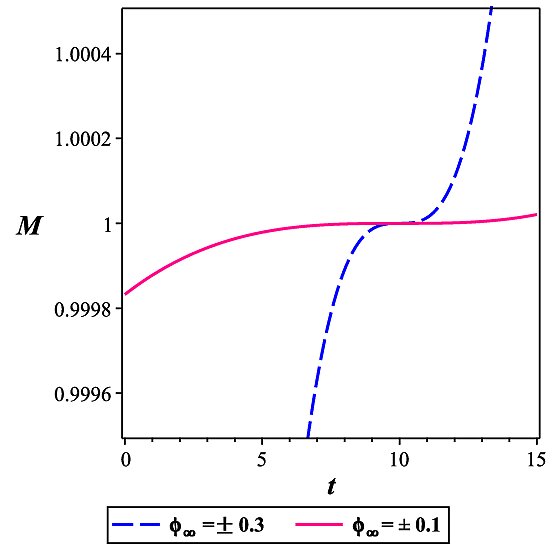}
	\end{center}
	\caption{\small {Time evolution of the black hole mass
			for different values of $\phi_{\infty}$, with quadratic potential.
			The left panel is corresponding to the case with $\phi_{\infty}=0.1\,\dot{\phi}_{\infty}$ and the right panel is corresponding
			to the case with $\phi_{\infty}=-0.1\,\dot{\phi}_{\infty}$.}}
	\label{fig5}
\end{figure}

\begin{figure}[]
	\begin{center}
		\includegraphics[scale=0.35]{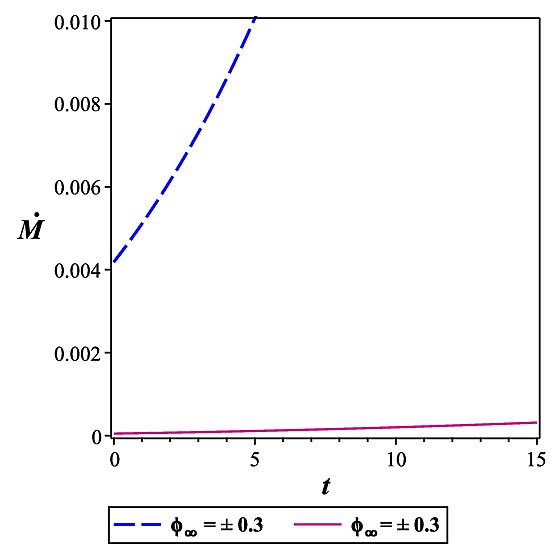}
		\includegraphics[scale=0.35]{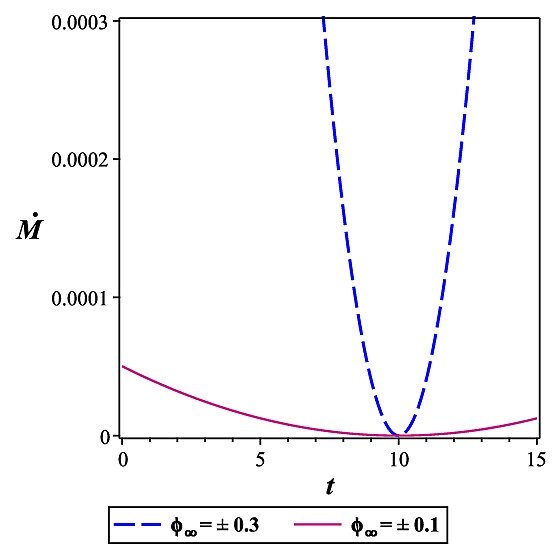}
	\end{center}
	\caption{\small {Time evolution of $\dot{M}$ for
			different values of $\phi_{\infty}$, with quadratic potential. The
			left panel is corresponding to the case with $\phi_{\infty}=0.1\,\dot{\phi}_{\infty}$ and the right panel is corresponding
			to the case with $\phi_{\infty}=-0.1\,\dot{\phi}_{\infty}$.}}
	\label{fig6}
\end{figure}

This means that, to the numerical study of the model, we should
consider the parameters' values satisfying the constraint
(\ref{eq30}). Here also, we perform a numerical analysis which is
shown in figure 4. As the linear case, we have considered
$\phi_{\infty}=0.1\,\dot{\phi}_{\infty}$ and
$\phi_{\infty}=-0.1\,\dot{\phi}_{\infty}$ and also adopted $M_{0}=1$
and $t_{0}=0$. As figure shows, with $\phi=0.1\,\dot{\phi}$, the
constraint (\ref{eq30}) implies some constraint on the parameter
$\phi$. However, with $\phi_{\infty}=-0.1\,\dot{\phi}_{\infty}$,
there is no constraint on $\phi_{\infty}$ and all considered values
of $\phi_{\infty}$ are viable. Now, we can study the evolution of
the black hole mass in the quadratic potential case. In this regard,
we analyze the time evolution of the black hole mass by using
equation (\ref{eq28}). The results are shown in figure 5. As figure
shows, for both cases with $\phi_{\infty}=0.1\,\dot{\phi}_{\infty}$
and $\phi_{\infty}=-0.1\,\dot{\phi}_{\infty}$ and with quadratic
potential, the black hole mass increases continually.
Correspondingly, the value of $\dot{M}$ in every time is positive.
This issue has been shown in figure 6. As a result, the accretion of
the tachyon field with quadratic potential into the black hole leads
to an increase in the black hole mass. Note that, with a quadratic
potential, we have $\rho+p<0$, leading to violation of both null and
weak energy conditions. This means that a tachyon field with
quadratic potential behaves like a phantom (despite it is not a
usual phantom field).

\subsection{Natural Potential}
With the natural potential $V\sim
\Lambda^{4}\big[1+\cos\big(\frac{\phi}{l}\big)\big]$~\cite{Fre90,Fre04,Fre14},
we find the function $f(t)$ as
\begin{eqnarray}
\label{eq31} f(t)=8\,\pi\,\Lambda^{4} \left[ 1+\cos \left( {\frac
	{\dot{\phi}_{_{\infty}}\, \left( t-t_{0} \right) +\phi_{_{\infty}}}{l}} \right) \right] \dot{\phi}^{2}_{_{\infty}} \left(
\dot{\phi}^{2}_{_{\infty}}+2 \right)\,,
\end{eqnarray}
and ${\cal{F}}$ as
\begin{eqnarray}
\label{eq32}
{\cal{F}}=8\,\pi\,\Lambda^{4}\,\dot{\phi}^{2}_{_{\infty}} \left(
\dot{\phi}^{2}_{_{\infty}}+2 \right) \Bigg[ t+\frac
{l}{\dot{\phi}_{_{\infty}}} \sin \left( {\frac
	{\dot{\phi}_{_{\infty}}\,t}{l}}-{\frac {\dot{\phi}_{_{\infty}}\,
		t_{0}-\phi_{_{\infty}}}{l}} \right)  \Bigg]\,.
\end{eqnarray}
Now, from equations (\ref{eq17}), (\ref{eq19}), (\ref{eq31}) and
(\ref{eq32}), we find the black hole mass and its time evolution in
the natural potential case as follows
\begin{eqnarray}
\label{eq33} M= M_{0}\Bigg[
-8\,\pi{\Lambda}^{4}\,\dot{\phi}^{2}_{_{\infty}} \left(
\dot{\phi}^{2}_{_{\infty}} +2 \right)  \bigg[ t+\frac
{l}{\dot{\phi}_{_{\infty}}} \sin \left( {\frac {t\,
		\dot{\phi}_{_{\infty}}}{l}}-{\frac {\dot{\phi}_{_{\infty}}\,
		t_{0}-\phi_{_{\infty}}}{l}} \right)  \bigg] M_{0}+1\Bigg]\,,
\end{eqnarray}

\begin{eqnarray}
\label{eq34}
\dot{M}=\Bigg[8\,\pi\,{\Lambda}^{4}\,\dot{\phi}^{2}_{_{\infty}}\left(\dot{\phi}^{2}_{_{\infty}}+2
\right) {
	M_{0}}^{2} \bigg( 1+\cos \left( {\frac {\dot{\phi}_{_{\infty}}\,
		\left( t- t_{0} \right) +\phi_{_{\infty}}}{l}} \right) \bigg)\Bigg]\hspace{1.5cm}\nonumber\\ \Bigg[ \bigg[
-8\,\pi{\Lambda}^{4}\,\dot{\phi}^{2}_{_{\infty}} \left(
\dot{\phi}^{2}_{_{\infty}}+2 \right)  \bigg( t+\frac {l}{
	\dot{\phi}_{_{\infty}}} \sin \left( {\frac {t\,\dot{\phi}{_{\infty}}}{l}}-{\frac
	{\dot{\phi}_{_{\infty}}\,t_{0}-\phi_{_{\infty}} }{l}} \right)
\bigg) M_{0}+1 \bigg] ^{2}\Bigg] \,.
\end{eqnarray}
For the natural potential case, equation (\ref{eq33}) gives the
following constraint on the model's parameters
\begin{eqnarray}
\label{eq35} 8\pi{\Lambda}^{4}\dot{\phi}^{2}_{_{\infty}} \left(
\dot{\phi}^{2}_{_{\infty}} +2 \right)  \Bigg[ t+\frac
{l}{\dot{\phi}_{_{\infty}}}\sin \left( {\frac {t
		\dot{\phi}_{_{\infty}}}{l}}-{\frac {\dot{\phi}_{_{\infty}}
		t_{0}-\phi_{_{\infty}}}{l}} \right)  \Bigg] M_{0}<1.
\end{eqnarray}

\begin{figure}[]
	\begin{center}
		\includegraphics[scale=0.5]{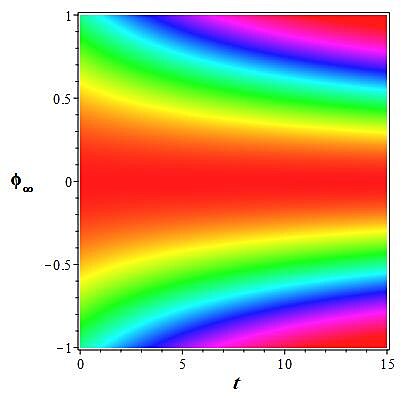}
		\includegraphics[scale=0.55]{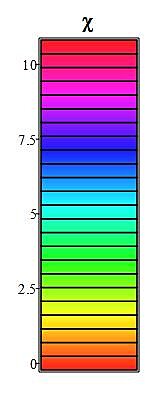}
		\includegraphics[scale=0.5]{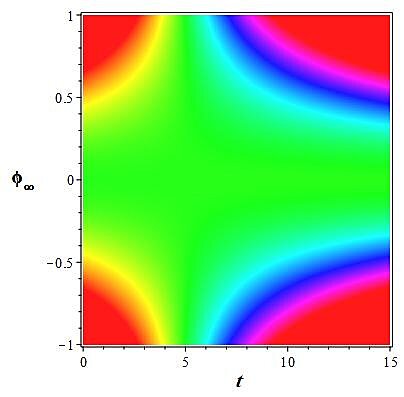}
		\includegraphics[scale=0.54]{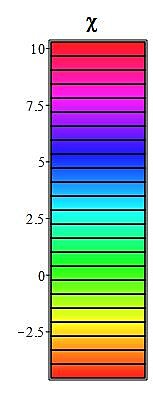}
	\end{center}
	\caption{\small {Presentation of the range of parameter
			$\phi_{\infty}$ and $t$, for the natural potential, satisfying the
			constraint (\ref{eq35}). The left panel is corresponding to the case
			with $\phi_{\infty}=0.1\,\dot{\phi}_{\infty}$ and the right panel is
			corresponding to the case with
			$\phi_{\infty}=-0.1\,\dot{\phi}_{\infty}$. Also, the parameter
			$\chi$ represents the left hand side of equation (\ref{eq35}).}}
	\label{fig7}
\end{figure}

\begin{figure}[]
	\begin{center}
		\includegraphics[scale=0.35]{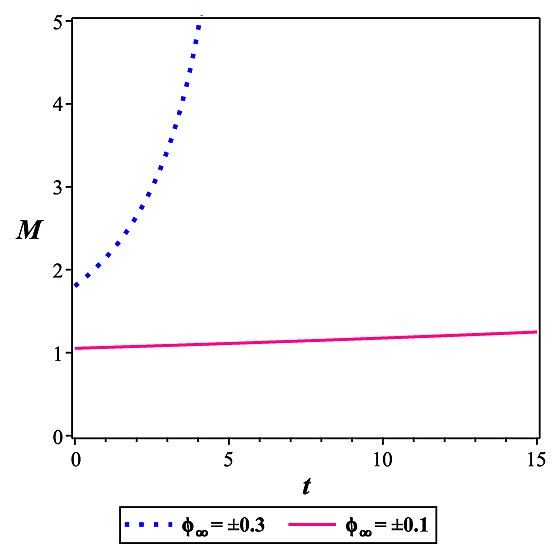}
		\includegraphics[scale=0.35]{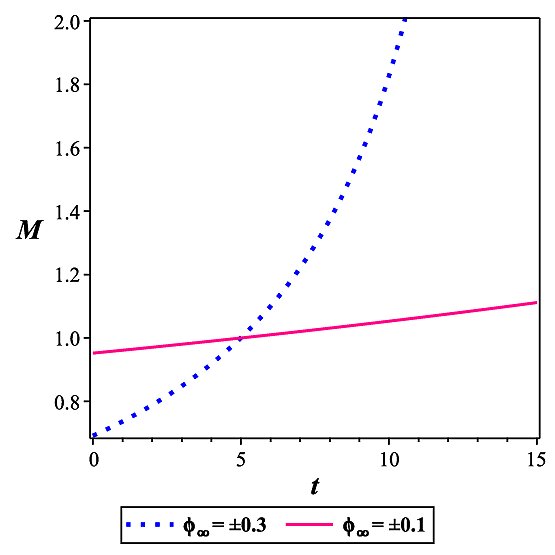}
	\end{center}
	\caption{\small {Time evolution of the black hole mass
			for different values of $\phi_{\infty}$, for the natural potential.
			The left panel is corresponding to the case with $\phi_{\infty}=0.1\,\dot{\phi}_{\infty}$ and the right panel is corresponding
			to the case with $\phi_{\infty}=-0.1\,\dot{\phi}_{\infty}$.}}
	\label{fig8}
\end{figure}

\begin{figure}[]
	\begin{center}
		\includegraphics[scale=0.35]{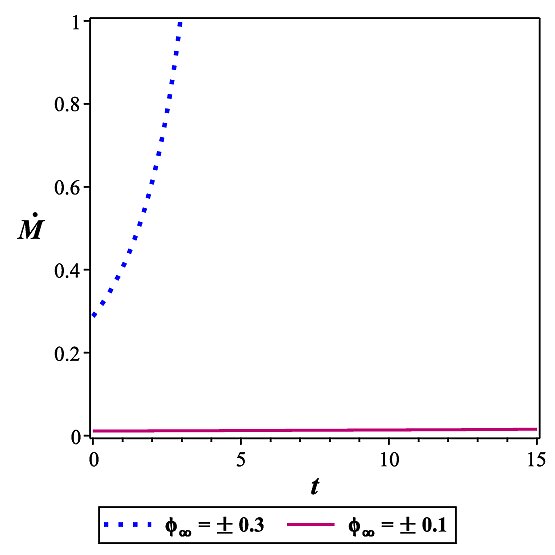}
		\includegraphics[scale=0.35]{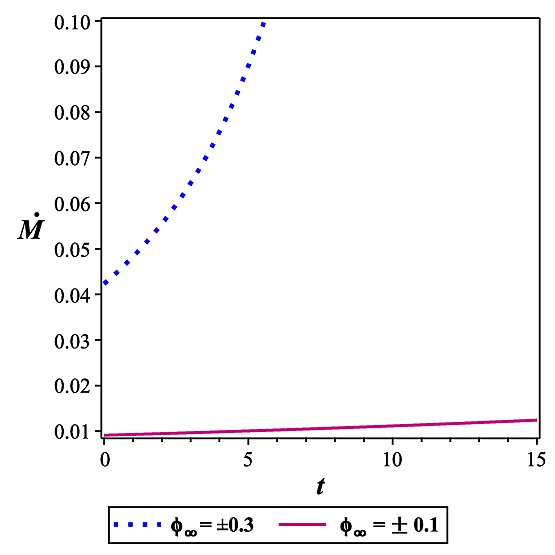}
	\end{center}
	\caption{\small {Time evolution of $\dot{M}$ for
			different values of $\phi_{\infty}$, for the natural potential. The
			left panel is corresponding to the case with $\phi_{\infty}=0.1\,\dot{\phi}_{\infty}$ and the right panel is corresponding
			to the case with $\phi_{\infty}=-0.1\,\dot{\phi}_{\infty}$.}}
	\label{fig9}
\end{figure}

The parameter space satisfying the constraint (\ref{eq35}) is shown
in figure 7, for two cases with
$\phi_{\infty}=0.1\,\dot{\phi}_{\infty}$ and
$\phi_{\infty}=-0.1\,\dot{\phi}_{\infty}$. To plot this figure, we
have considered $\Lambda=1$, $l=1$, $M_{0}=1$ and $t_{0}=0$. As the
figure shows, for both cases, there are some values of the
parameters which satisfy the constraint (\ref{eq35}) and can be used
in numerical analysis of $M$ and $\dot{M}$. Here, we consider
$\phi_{\infty}=-0.3$ , $-0.1$, $0.1$ and $0.3$. With these adopted
values of $\phi_{\infty}$, we study the evolution of the black hole
mass versus time for both cases with
$\phi_{\infty}=0.1\,\dot{\phi}_{\infty}$ and
$\phi_{\infty}=-0.1\,\dot{\phi}_{\infty}$. The results are shown in
figure 8. This figure shows that, as time goes, the mass of the
black hole increases. This means that with natural potential we
always have the accretion of the mass into the black hole and there
is no evaporation or decreasing of the black hole mass. This result
is verified by figure 9, where we have plotted the time evolution of
$\dot{M}$. When we adopt a natural potential for the tachyon field,
we find $\rho+p<0$. Therefore, in this case also, both null and weak
energy conditions are violated. In this regard, it seems that a
tachyon field with a natural potential behaves like a phantom.

\subsection{Hilltop Potential}
With the hilltop potential which is defined as $V\sim
V_{0}\left[1-\left(\frac{\phi}{\mu}\right)^{q}\right]$~\cite{Fre90,Fre04,Fre14},
we find the following expression for $f(t)$
\begin{eqnarray}
\label{eq36} f(t)=16\,\pi\,V_{0} \left[ 1- \left( {\frac
	{\dot{\phi}_{_{\infty}}\, \left( t- t_{0} \right) +\phi_{\infty}}{\mu}} \right)^{q} \right] \dot{\phi}^{2}_{_{\infty}} \left(
1+\frac{1}{2}\, \dot{\phi}^{2}_{_{\infty}} \right) .
\end{eqnarray}
Also, the hilltop potential gives ${\cal{F}}$ as
\begin{eqnarray}
\label{eq37} {\cal{F}}=8\,\pi\,V_{0}\dot{\phi}^{2}_{_{\infty}}
\left(\dot{\phi}^{2}_{_{\infty}}+2 \right) \Bigg( t-\frac
{l}{\dot{\phi}_{_{\infty}}\, \left( 1+q \right) } \left( {\frac {
		\dot{\phi}_{_{\infty}}\,t}{\mu}}+{\frac
	{-\dot{\phi}_{_{\infty}}\, t_{0}+\phi_{\infty}}{\mu}} \right)
^{1+q} \Bigg) \,.
\end{eqnarray}
By having the above equations and using equations (\ref{eq17}) and
(\ref{eq19}), we find the black hole mass as follows
\begin{eqnarray}
\label{eq38} M= M_{0}\Bigg[ \Bigg( 1-8\pi
V_{0}\dot{\phi}^{2}_{_{\infty}} \left( \dot{\phi}^{2}_{_{\infty}} +2
\right)  \bigg( t-\frac {\mu}{\dot{\phi}_{_{\infty}}\,
	\left( 1+q \right) }\nonumber\\ \left( {\frac {\dot{\phi}_{_{\infty}}\,t}{\mu}}-
{\frac {\dot{\phi}_{_{\infty}}\,t_{0}-\phi_{\infty}}{\mu}} \right) ^{1+q} \bigg)  M_{0}
\Bigg)\Bigg] \,,
\end{eqnarray}
and its time evolution as
\begin{eqnarray}
\label{eq39} \dot{M}=\Bigg[16\pi V_{0}\dot{\phi}^{2}_{_{\infty}}
\left( 1+\frac{1}{2}\dot{\phi}^{2}_{_{\infty}} \right)
M_{0}^{2} \bigg( 1-\left( {\frac {\dot{\phi}_{_{\infty}}\ \left(
		t- t_{0} \right) +\phi_{\infty}}{\mu}} \right) ^{q} \bigg)\Bigg]\hspace{1.5cm}\nonumber\\ \Bigg[
\Bigg( 1-8\pi V_{0}\dot{\phi}^{2}_{_{\infty}} \left(
\dot{\phi}^{2}_{_{\infty}}+2 \right) \bigg( t- \frac
{\mu}{\dot{\phi}_{_{\infty}}\left( 1+q \right) }  \left( {\frac
	{\dot{\phi}^{2}_{_{\infty}}t}{\mu}}- {\frac {\dot{\phi}_{_{\infty}} t_{0}-\phi_{\infty}}{l}} \right) ^{1+q} \bigg) M_{0}
\Bigg)^{2} \Bigg] .\nonumber\\
\end{eqnarray}
For the hilltop potential case, equation (\ref{eq38}) gives the
following constraint on the model's parameters
\begin{eqnarray}
\label{eq40} 8\pi V_{0}\dot{\phi}^{2}_{_{\infty}} \left(
\dot{\phi}^{2}_{_{\infty}} +2 \right)  \Bigg( t-\frac
{\mu}{\dot{\phi}_{_{\infty}} \left( 1+q \right) }  \left( {\frac
	{\dot{\phi}_{_{\infty}}t}{\mu}}-{\frac {\dot{\phi}_{_{\infty}}t_{0}-\phi_{\infty}}{\mu}} \right) ^{1+q} \Bigg)
M_{0}<1\,.
\end{eqnarray}

\begin{figure}[]
	\begin{center}
		\includegraphics[scale=0.5]{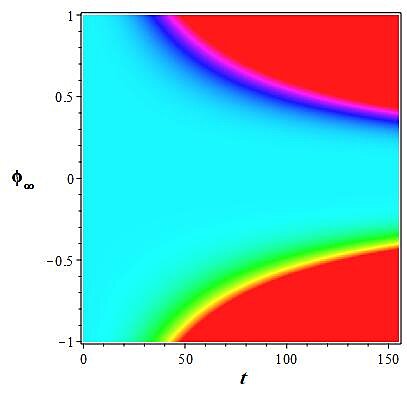}
		\includegraphics[scale=0.54]{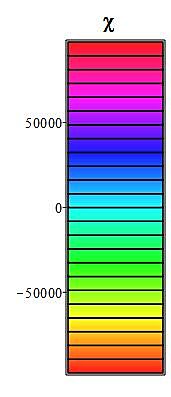}
		\includegraphics[scale=0.5]{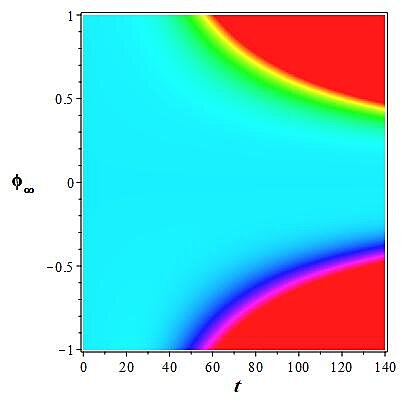}
		\includegraphics[scale=0.53]{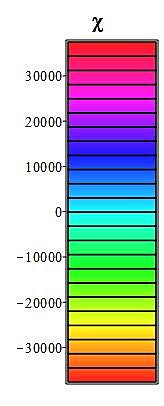}
	\end{center}
	\caption{\small {Presentation of the range of parameter
			$\phi_{\infty}$ and $t$, with hilltop potential, satisfying the
			constraint (\ref{eq40}). The left panel is corresponding to the case
			with $\phi_{\infty}=0.1\,\dot{\phi}_{\infty}$ and the right panel
			is corresponding to the case with $\phi_{\infty}=-0.1\,\dot{\phi}_{\infty}$. Also, the parameter $\chi$
			represents the left hand side of equation (\ref{eq40}).}}
	\label{fig10}
\end{figure}

\begin{figure}[]
	\begin{center}
		\includegraphics[scale=0.34]{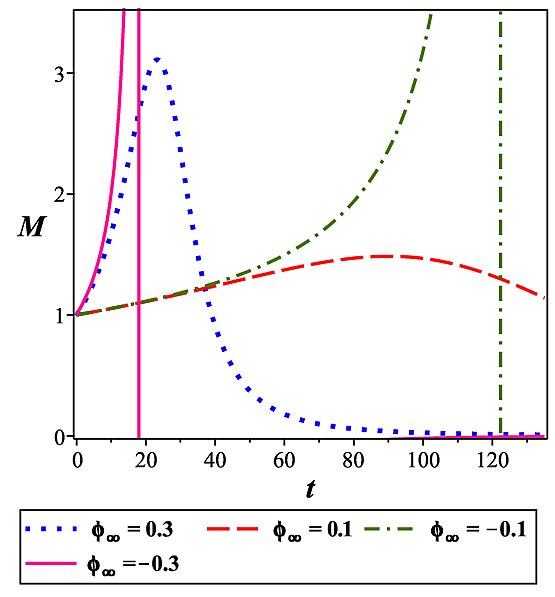}
		\includegraphics[scale=0.35]{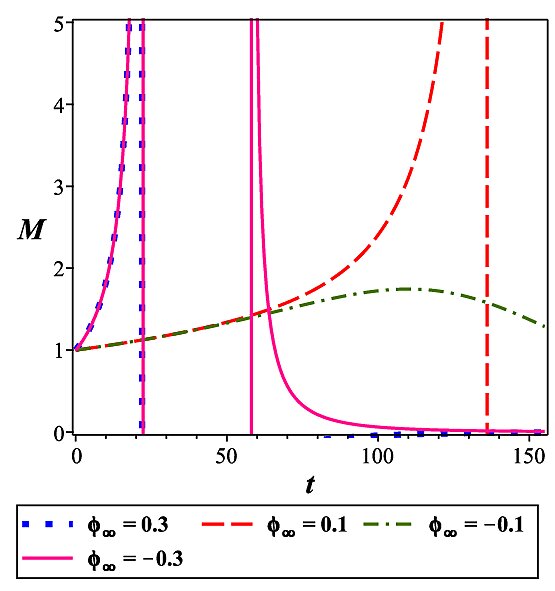}
\end{center}
	\caption{\small {Time evolution of the black hole mass
			for different values of $\phi_{\infty}$, with hilltop potential.
			The left panel is corresponding to the case with $\phi_{\infty}=0.1\,\dot{\phi}_{\infty}$
			and the right panel is corresponding
			to the case with $\phi_{\infty}=-0.1\,\dot{\phi}_{\infty}$.}}
	\label{fig11}
\end{figure}

\begin{figure}[]
	\begin{center}
		\includegraphics[scale=0.345]{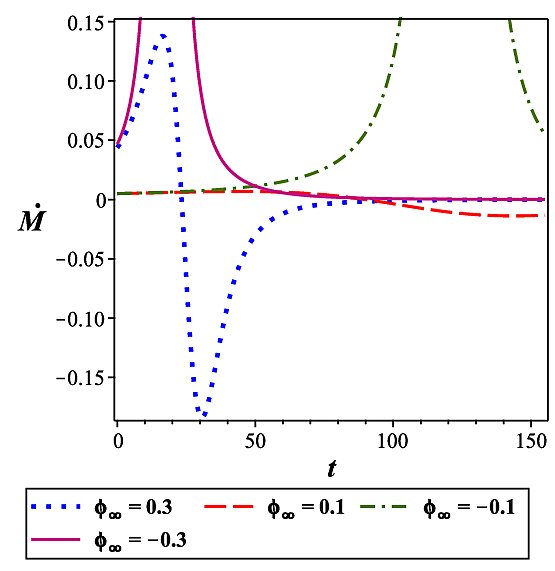}
		\includegraphics[scale=0.35]{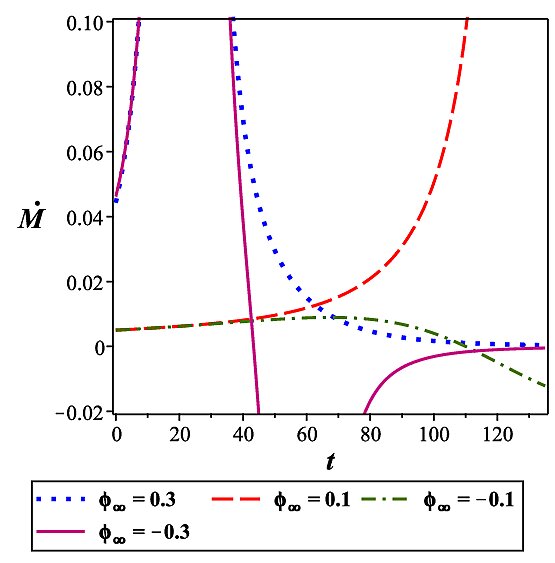}
	\end{center}
	\caption{\small {Time evolution of $\dot{M}$ for
			different values of $\phi_{\infty}$, with hilltop potential. The
			left panel is corresponding to the case with $\phi_{\infty}=0.1\,\dot{\phi}_{\infty}$ and the right panel is corresponding
			to the case with $\phi_{\infty}=-0.1\,\dot{\phi}_{\infty}$.}}
	\label{fig11}
\end{figure}

The values of the model's parameter which is used to perform
numerical analysis on the model with hilltop potential should
satisfy the constraint (\ref{eq40}). The acceptable ranges of the
parameters are shown in figure 10, for two cases with
$\phi_{\infty}=0.1\,\dot{\phi}_{\infty}$ and
$\phi_{\infty}=-0.1\,\dot{\phi}_{\infty}$. In this figure, the
parameter $\chi$ is the left-hand side of the constraint
(\ref{eq40}). The viable regions of $\phi_{\infty}$ and $t$ are
those that lead to $\chi<1$. By using these viable values, we study
the evolution of the black hole mass versus time numerically. The
results are shown in figure 11. Our numerical analysis shows that,
in the case with $\phi_{\infty}=0.1\,\dot{\phi}_{\infty}$ and for
$\phi_{\infty}=0.1$ and $\phi_{\infty}=0.3$, the presence of the
tachyon field causes the black hole mass first increases and then
decreases by time. Therefore, with hilltop potential, it is possible
to have a decreasing black hole mass without considering the Hawking
radiation. In the case with
$\phi_{\infty}=-0.1\,\dot{\phi}_{\infty}$, if we consider
$\phi_{\infty}=-0.1$ and $\phi_{\infty}=-0.3$, it is possible to
have decrease in the mass of the black hole. However, for
$\phi_{\infty}=-0.3$, the evolution of the black hole mass follows a
strange pattern and is not favor. The time evolution of $\dot{M}$
for both cases with $\phi_{\infty}=0.1\,\dot{\phi}_{\infty}$ and
$\phi_{\infty}=-0.1\,\dot{\phi}_{\infty}$ is shown in figure 12.
This figure also confirms that, for
$\phi_{\infty}=0.1\,\dot{\phi}_{\infty}$ ,$\phi_{\infty}=0.1$ and
$\phi_{\infty}=0.3$ and also for
$\phi_{\infty}=-0.1\,\dot{\phi}_{\infty}$, $\phi_{\infty}=-0.1$ and
$\phi_{\infty}=-0.3$, the mass of the black hole decreases even
there is no Hawking radiation.

Here also, in the ranges of the parameters space leading to black
hole mass decrease, we have $\rho+p>0$. Therefore, in these ranges,
the null energy condition is not violated. This means that, in these
ranges, the tachyon field is not a phantom. As the linear potential
case, the mass decrease of the black hole occurs due to the negative
energy density of the tachyon field.

\section{Summary} \label{sec4}

In this paper, we have studied the accretion of the black hole mass
in the presence of the tachyon field. Given that the previous works
on the black hole mass accretion were with canonical scalar fields,
in both the minimally and non-minimally coupled cases, we have
considered a non-canonical scalar field on this issue. As regards
the energy-momentum tensor of the tachyon field has non-diagonal
components, the potential of the field is involved in the equations
corresponding to the mass and its time evolution. In this regard,
there is no need for the non-minimal coupling between the tachyon
field and gravity. We have used four types of potential as linear,
quadratic, natural, and hilltop potentials. To study the black hole
mass, we have used the fact that some interesting cosmological
solutions are corresponding to the negative potentials. In this
regard, we have used the parameter space leading to both the
negative and positive potentials. We have restricted the values of
the model's parameters by implying the constraint $M\geq 0$, where
$M$ is the black hole mass. To show those values of the model's
parameters satisfying the constraint $M\geq 0$, we have plotted some
figures which gave us the viable ranges of the parameters. We have
used these viable ranges to study the time evolution of the
parameters $M$ and $\dot{M}$. By numerical analysis of $M$ and
$\dot{M}$, we were wonder whether black hole mass decreases or not.
Our numerical study has shown that if we consider the linear
potential, for the cases with $\phi_{\infty}=\pm
0.1\,\dot{\phi}_{\infty}$, $\phi_{\infty}=0.1$ and
$\phi_{\infty}=0.3$ we have accretion of the mass into the black
hole. However, with $\phi_{\infty}=\pm 0.1\,\dot{\phi}_{\infty}$,
$\phi_{\infty}=-0.1$ and $\phi_{\infty}=-0.3$, the black hole
decreases without considering the Hawking radiation.

Another interesting point is that the tachyon field with linear and
hilltop potentials, in some ranges of the parameter space leading to
the black hole mass decrease, satisfies the null energy condition.
In the cases with quadratic and natural potentials, the null energy
condition is violated and the tachyon field has ``phantom-like"
behavior.
\\

{\bf Acknowledgement}\\
I thank the referee for the very insightful comments that have improved the quality of the paper considerably.\\

%\begin{thebibliography}{000} %for 3 digits
%\begin{thebibliography}{00}  %for 2 digits

\end{document}